\documentclass[10pt,showpacs,letterpaper,aps,twocolumn,nofootinbib,prx]{revtex4-1} 

\usepackage{amsthm,amsmath,amssymb,graphicx,comment} 
\usepackage{color}
\usepackage{bbold}
\usepackage[none]{hyphenat}
\usepackage{hyperref}

\newcommand{\beq}{\begin{equation}}
\newcommand{\eeq}{\end{equation}}
\newcommand{\bea}{\begin{align}}
\newcommand{\eea}{\end{align}}

\begin{document}

\title{The ontological identity of empirical indiscernibles: Leibniz's methodological principle and its significance in the work of Einstein}

\author{Robert W. Spekkens}
\affiliation{Perimeter Institute for Theoretical Physics, 31 Caroline St. N, Waterloo, Ontario, N2L 2Y5, Canada}

\date{August 28, 2019}                                           

\begin{abstract} 
This article explores the following methodological principle for theory construction in physics: if an ontological theory predicts two scenarios that are ontologically distinct but empirically indiscernible, then this theory should be rejected and replaced by one relative to which the scenarios are ontologically the same.  I defend the thesis that this methodological principle was first articulated by Leibniz as a version of his principle of the identity of indiscernibles, and that it was applied repeatedly to great effect by Einstein in his development of the special and general theories of relativity.  I argue for an interpretation of the principle as an inference to the best explanation, defend it against some criticisms, discuss its potential applications in modern physics, and explain how it provides an attractive middle ground in the debate between empiricist and realist philosophies of science.

\end{abstract}

\maketitle

\section{Introduction}


This article concerns the version of Leibniz's principle of the identity of indiscernible that is found in the Leibniz-Clarke correspondence~\cite{LeibnizClarke}.  The following quotation, point 6 of the 4th letter of the correspondence, highlights the usefulness of the principle in the debate regarding whether motion is absolute or relative:
\begin{quote}
To suppose two things indiscernible, is to suppose the same thing under two names. And therefore to suppose that the universe could have had at first another position of time and place, than that which it actually had; and yet that all the parts of the universe should have had the same situation among themselves, as that which they actually had; such a supposition, I say, is an impossible fiction.
\end{quote}


I take Leibniz to be using his principle of the identity of indiscernibles here as a normative maxim, a methodological principle that guides the construction of physical theories.  Specifically, I take the maxim at play to be that the {\em ontological} account of two hypothetical scenarios or phenomena ought to be the same whenever the {\em empirical} (or equivalently {\em operational}) account of these phenomena is the same. 


More explicitly, the principle that I wish to endorse,
and to which I want to claim Leibniz as the progenitor, is as follows:
\begin{quote} 
 {\bf The Leibnizian methodological principle}\\
If an ontological theory implies the existence of two scenarios that are empirically indistinguishable in principle but ontologically distinct 
(where both the indistinguishability and distinctness are evaluated
 by the lights of the theory in question),
then the ontological theory should be rejected and replaced with one relative to which the two scenarios are ontologically identical.
\end{quote}

Heuristically, the principle asserts the {\em ontological} identity of {\em empirical} indiscernibles. 
In previous work discussing the principle~\cite{spekkens2015paradigm}, I attributed it to Einstein (for reasons that will become apparent), but I shall here argue that it should be attributed to Leibniz.

 It is reasonable to wonder whether adding the ``ontological'' and ``empirical'' qualifiers to the notions of ``identity'' and ``indiscernibility'' in   Leibniz's principle of the identity of indiscernibles, 
  and casting it as a methodological principle for theory construction
 constitutes a significant departure from what Leibniz had in mind.  In fact, the Leibniz-Clarke correspondence supports the thesis that Leibniz endorsed this reading,
   at least in the context of the debate on the nature of space and motion.\footnote{Sklar~\cite{sklar1977space} has argued that one should read Leibniz's principle of the identity of indiscernibles as ontological identity given indiscernibility ``relative to all qualitative properties''.
 However, it is quite unclear what is meant by ``qualitative property''.  Furthermore, I see no evidence in Leibniz's discussion of the absolute-relative debate that he had in mind indiscernibility relative to some distinguished type of property, and much evidence that he had in mind {\em empirical} indistinguishability. 
  Indeed, Sklar also ultimately concludes that the only sensible notion of indiscernibility for Leibniz's argument is the empirical one.  It seems to me that this is in fact the only reading that is supported by the textual evidence.
    }


For instance, in point 52 of the 5th letter, Leibniz recapitulates his argument to Clarke as follows: 
\begin{quote}
'tis unreasonable it [the universe] should have any motion otherwise than as its parts change their situation among themselves; because such a motion would produce no change that could be observed, and would be without design. 
\end{quote}
He then paraphrases Clarke's prior response to this argument as follows:
\begin{quote}
the reality of motion does not depend upon being observed; and that a ship may go forward, and yet a man, who is in the ship, may not perceive it.
\end{quote}
To which he provides the following rebuttal: 
\begin{quote}
I answer, motion does not indeed depend upon being observed; but it does depend upon being possible to be observed. [...] when there is no change that can be observed, there is no change at all.
\end{quote}
His conclusion in this argument, that it is unreasonable to assign reality to the distinction between the universe being globally in motion and being globally at rest (as opposed to the universe undergoing relative motions among its parts), i.e., his denial of the ``reality of motion'' in this case, is an affirmation of the ontological identity of the two imagined scenarios.  Meanwhile, the latter two quotes make it clear that his {\em justification} for this conclusion is 
explicitly 
the   {\em empirical} indiscernibility of the two imagined scenarios (that is, the impossibility of discernment by observation), and in particular their {\em in-principle} empirical indiscernibility, rather than a parochial observer-dependent or technology-dependent species of empirical indiscernibility.



I now turn to arguing for the credentials of this principle in physics.  
Despite the fact that physicists do not explicitly champion this principle---indeed, they do not seem to even know about it---I will argue that Einstein made use of it at several
critical junctures in his development of the special and general theories of relativity.

\section{The significance of the Leibnizian methodological principle in Einstein's work}

\subsection{The induction experiment} 

\begin{figure}[h] 
\centering
\includegraphics[width=0.45\textwidth]{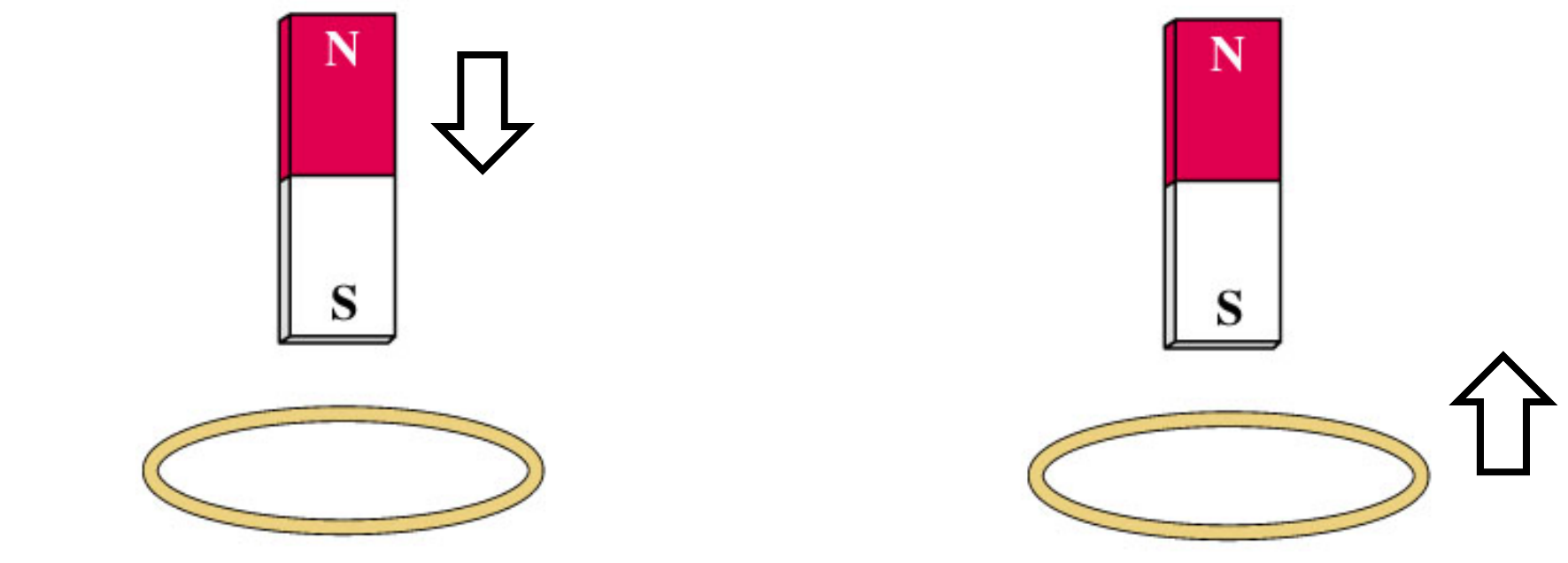}
\caption{The two scenarios considered by Einstein in his discussion of the induction experiment.} 
\label{InductionExpt}
\end{figure}

Consider the introductory paragraph to Einstein's 1905 paper ``On the electrodynamics of moving bodies''\cite{einstein1905electrodynamics} which introduced the special theory of relativity:
\begin{quote}
It is known that Maxwell's electrodynamics---as usually understood at the present time---when applied to moving bodies, leads to asymmetries which do not appear to be inherent in the phenomena. Take, for example, the reciprocal electrodynamic action of a magnet and a conductor. The observable phenomenon here depends only on the relative motion of the conductor and the magnet, whereas the customary view draws a sharp distinction between the two cases in which either the one or the other of these bodies is in motion.  For if the magnet is in motion and the conductor at rest, there arises in the neighbourhood of the magnet an electric field with a certain definite energy, producing a current at the places where parts of the conductor are situated. But if the magnet is stationary and the conductor in motion, no electric field arises in the neighbourhood of the magnet. In the conductor, however, we find an electromotive force, to which in itself there is no corresponding energy, but which gives rise---assuming equality of relative motion in the two cases discussed---to electric currents of the same path and intensity as those produced by the electric forces in the former case.
\end{quote}
Einstein is here emphasizing that relative to the understanding of electrodynamics that prevailed at the time, scenarios involving different motions {\em relative to the ether}, such as the  two scenarios depicted in Fig~\ref{InductionExpt}, were considered to be physically distinct and yet generated precisely the same observable phenomenon.  
Specifically, in the scenario where the conducting loop is at rest with respect to the ether while the bar magnet is moving relative to it, the static charges in the loop feel a time-varying magnetic field, hence an electric field, which induces a current. By contrast, in the scenario where the conducting loop is moving with respect to the ether while the bar magnet is stationary relative to it, the charges in the loop are in motion through a static magnetic field, and the associated Lorentz force induces a current.
 As Einstein emphasizes, if the relative motion is the same in the two scenarios, the induced current in the loop is identical; the observable phenomenon (the current) depends {\em only} on the relative motion. 

In summary, Einstein has shown that, for the understanding of electrodynamics that prevailed at the time, there are two physically distinct scenarios that are completely indistinguishable observationally.
The fact that Einstein takes this to be unacceptable and uses the induction experiment to motivate the development of a physical model devoid of a notion of ether or absolute rest, and wherein only the relative motion of objects is physically significant (which is what the special theory of relativity ultimately achieves) is an instance of his endorsement of the Leibnizian methodological principle.

 
 
\subsection{The equivalence principle}

In his account of the equivalence principle, Einstein considered
 the following two physical scenarios which are understood to be physically distinct according to Newtonian theory:
 (i) being at rest in a uniform gravitational field that induces acceleration $a$ in the direction ${-}\hat{n}$, or (ii) accelerating with acceleration $a$ in the direction ${+}\hat{n}$ through a space that is free of gravitational fields.
  Einstein noted that these two scenarios are observationally indistinguishable and concluded that they should be considered physically equivalent as well.  As such, it is another clear example of the application of Leibniz's methodological principle.

Another version of the principle considers the pair of scenarios to be: (i) freely falling in a uniform gravitational field,
  and (ii) moving at fixed velocity 
  through a space that is free of gravitational fields.  Einstein's conclusion about these is the same: observational indistinguishability implies physical equivalence. 

\begin{figure}[h] 
\centering
\includegraphics[width=0.45\textwidth]{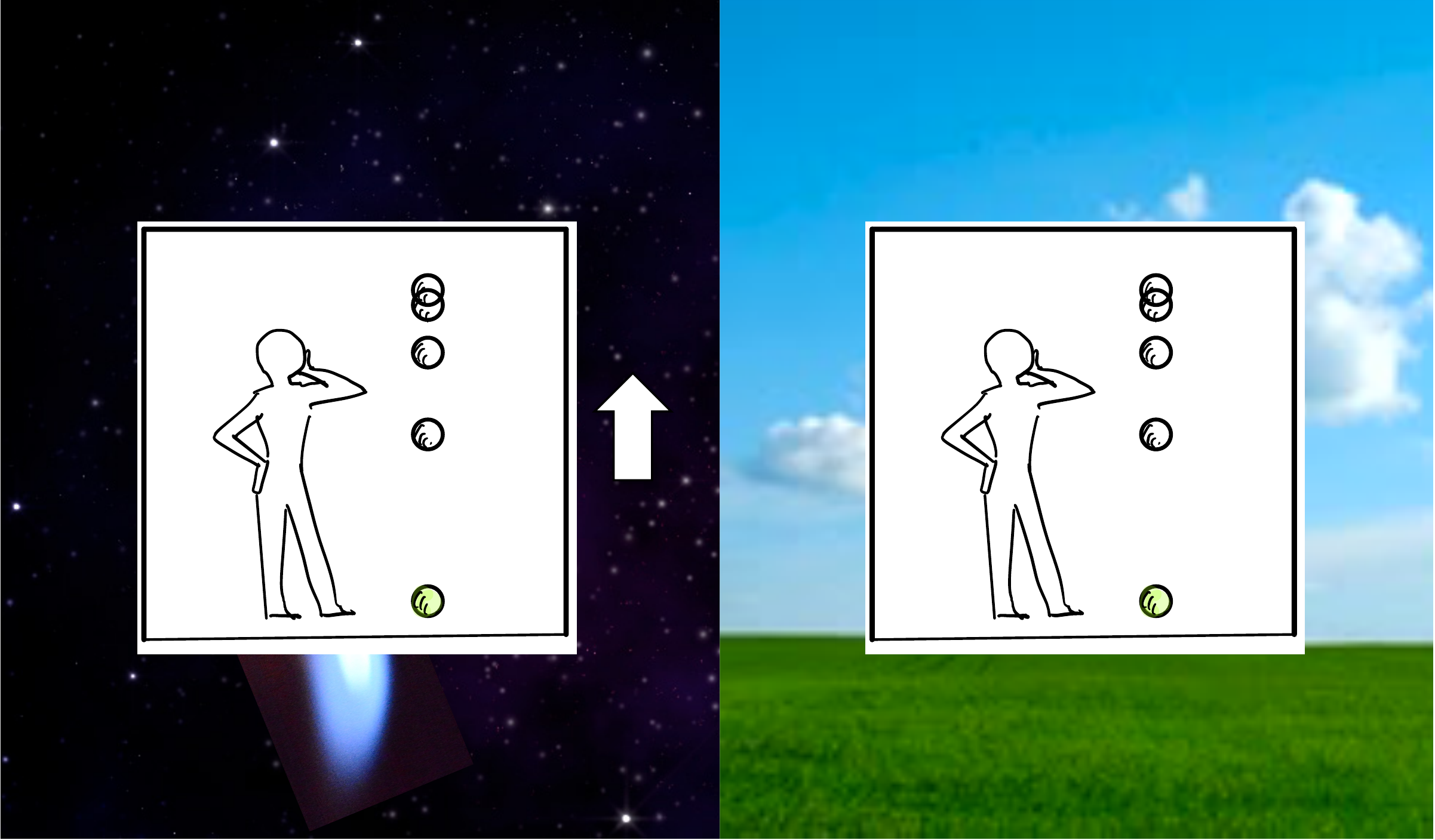}
\caption{The two scenarios considered by Einstein in his discussion of the equivalence principle.} 
\label{SEP}
\end{figure}

The observational equivalence of scenarios (i) and (ii) (for either version of the equivalence principle just articulated) is most often explained these days
  by invoking a thought experiment involving an observer inside an elevator who is implementing experiments and observing the outcomes, and by noting that nothing that she observes therein will allow her to deduce whether the elevator is at rest in a uniform gravitational field or whether it is accelerating through free space (alternatively, whether the elevator is freely falling in a uniform gravitational field or whether it is moving at constant velocity through free space).  This is depicted in Fig.~\ref{SEP}.

To defend the claim that Einstein's argument did indeed conform to the Leibnizian methodological principle, we consider some textual evidence from his 1911 article ``On the influence of gravitation on the propagation of light''~\cite{einstein1911influence}.  
Einstein imagines a system of coordinates, $K$, that is stationary with respect to a uniform gravitational field inducing acceleration $a$ in some direction, and another, $K'$, moving with uniform acceleration $a$ in the opposite direction.

He notes that the form of the equations of motion of a material object relative to $K$ and relative to $K'$ are precisely the same. He further notes that while it is obvious in $K'$ that all material objects will experience the same acceleration, in $K$, on the other hand, the fact that they do is an instance of what he refers to as ``Galileo's principle'', that is, the equality of gravitational and inertial mass.
  Einstein then notes:
\begin{quote}
This experience, of the equal falling of all bodies in the gravitational field, is one of the most universal which the observation of nature has yielded, but in spite of that the law has not found any place in the foundations of our edifice of the physical universe.
\end{quote}
He then immediately suggests how the situation can be remedied:
\begin{quote}
We arrive at a very satisfactory interpretation of this law of experience, if we assume that the systems $K$ and $K'$ are physically exactly equivalent, that is, if we assume that we may just as well regard the system $K$ as being in a space free from gravitational fields, if we then regard $K$ as uniformly accelerated.  This assumption of exact physical equivalence makes it impossible for us to speak of the absolute acceleration of the system of reference, just as the usual theory of relativity forbids us to talk of the absolute velocity of a system; and it makes the equal falling of all bodies in a gravitational field seem a matter of course.
\end{quote}
Einstein is here pointing out that one secures a particularly satisfying account of the observational equivalence of the two scenarios (indeed, one that he describes as ``a matter of course'') by following the dictates of the Leibnizian methodological principles and assuming their physical equivalence.

That it is indeed the {\em observational} equivalence of the two scenarios which serves as the antecedent of his argument is clear from the fact that he refers to the equal falling of all bodies in a gravitational field 
 as a ``law of experience'' and something that ``the observation of nature has yielded''.
  The fact that the conclusion he seeks to draw is the {\em ontological} equivalence of the two scenarios
is plain from the fact that he speaks explicitly of their ``exact physical equivalence''\footnote{This same terminology is used elsewhere when he is summarizing the equivalence principle, for instance, in Ref.\cite{einstein1907relativity},
``we [...] assume the complete physical equivalence of a gravitational field and a corresponding acceleration of the reference system.''}. Furthermore, note the similarity between Leibniz's notion of two scenarios being ``the same thing by two names'' and Einstein's turn of phrase ``we may just as well regard the system $K$ as being in a space free from gravitational fields''.

We read all of this as Einstein arguing {\em against} the ontological picture of reality assumed prior to general relativity, wherein accelerations and gravitational field strengths were considered absolute notions and wherein the distinction between scenarios involving one or the other were considered meaningful.  He argues for supplanting this picture with a new ontological scheme wherein the distinction is abolished from the conceptual apparatus of the theory (so that it becomes ``impossible for us to speak of the absolute acceleration of the system of reference'').  Indeed, this abolishment of the distinction is ultimately achieved in his general theory of relativity.  Most importantly for our purposes here, in making his argument, Einstein explicitly appeals to the intuition underlying Leibniz's methodological principle. 





The notion that it is the {\em equivalence principle} (rather than some other principle) that served as the primary conceptual foundation on which the theory of general relativity was built
 is supported by Einstein's own reminiscences concerning its origin (recounted in a talk he gave in Kyoto in 1922~\cite{einstein2005constructed}):
\begin{quote}
The breakthrough came suddenly one day. I was sitting on a chair in my patent office in Bern. Suddenly the thought struck me: If a man falls freely, he would not feel his own weight. I was taken aback. This simple thought experiment made a deep impression on me. This led me to the theory of gravity. 
\end{quote}
Insofar as the equivalence principle is an instance of the Leibnizian methodological principle,
this establishes the importance of the latter in the conceptual underpinnings of general relativity. 


\subsection{The hole argument and the point coincidence argument}

In the years leading up to 1915, Einstein struggled to complete the general theory of relativity.  A key issue was whether he should insist on the principle of general covariance, that ``Laws of Nature are expressed by means of equations which are valid for all co-ordinate systems, that is, which are covariant for all possible transformations''~\cite{einstein1916foundation}.  The historical account that I shall relay---which has emerged as the standard one among historians and philosophers of science---originated in the work of Stachel and of Norton~\cite{stachel1985einstein, norton1984einstein} (see also Refs.~\cite{stachel2014hole,NortonSEP}).

In 1913, Einstein devised an argument  which for two years he took (mistakenly) to be evidence against general covariance.
This is the famous ``hole argument'', presented by Einstein in Sec. 12 of ``On the Foundations of the Generalized Theory of Relativity and the Theory of Gravitation"~\cite{einstein1914foundations}. 
Einstein considers a spacetime containing matter but also a region $\Sigma$ that is free of matter---which he termed the `hole'.  He then notes that one can define diffeomorphisms (i.e., nonsingular coordinate transformations) which act as identity outside of the hole but act nontrivially within the hole.  It follows that one can identify \emph{different} solutions (relative to a single coordinate system $K$) for the field inside the hole---he denotes them $G(x)$ and $G'(x)$---for a \emph{fixed} distribution of matter outside the hole. 
Einstein puts it as follows:
\begin{quote}
[...] assume the differential equations of the gravitational field to be everywhere covariant [...]
There are then two different solutions $G(x)$ and $G'(x)$ relative to $K$, even though the solutions coincide on the boundary of the domain $\Sigma$.  In other words, {\em the course of events in this domain cannot be determined uniquely by general-covariant differential equations}.
\end{quote}
Unwilling to abide nonuniqueness of the course of events, Einstein concludes that one must introduce restrictions on the choice of coordinate system and hence retreat from demanding general covariance.



Einstein only finds the flaw in his argument two years later and he
comes to the conclusion that the two different solutions {\em do not} represent physically distinct states of affairs and therefore that there is no challenge to the uniqueness of the {\em physical course of events} in the hole.
He reports his new conclusion
 for the first time in a letter to Ehrenfest of December 26, 1915~\cite{EinsteinCollectedPapers}.  The letter is explicitly a retraction of his former argument.~\footnote{As Einstein himself quips in the letter: ``Einstein has it easy.  Every year he retracts what he wrote in the preceding year; now the sorry business falls to me of justifying my latest retraction.''} It proceeds as follows:
\begin{quote}
A contradiction to the uniqueness of the events does not follow at all from the fact that both systems $G(x)$ and $G'(x)$, related to the same frame of reference, satisfy conditions of the gravitational field. [...]\\
In place of Sec.~12, the following consideration must appear.  
Whatever is physically real in events in the universe (as opposed to that which is dependent on the choice of a reference system) consist in {\em spatio-temporal coincidences}) and in nothing else!  For example, the intersection points of two world lines are real, or the statement that they do {\em not} intersect each other.  Therefore, those statements relating to the physically real do not lose validity from the absence of a (unique) coordinate transformation.  When two systems in the $g_{\mu \nu}$'s (or generally, the variables used to describe the world) are constituted in such a way that the second can be obtained from the first by mere space-time transformation, then they are entirely equivalent.  This is because they have in common all the spatio-temporal point coincidences, that is, all the observables. 

This consideration shows simultaneously how natural the requirement of general covariance is.
\end{quote}

This has come to be known as Einstein's \emph{point-coincidence argument}.

I now draw out explicitly how this constitutes an endorsement of the Leibnizian methodological principle. 
The hole argument stipulates two solutions that are represented differently in the mathematical formalism; the question arises, however, of whether or not they should be considered to be {\em physically} distinct.  

 
 Einstein ultimately answers the question in the negative by appealing to facts about what is {\em observable}.  Specifically, he argues that because the only aspects of a solution that are observable are space-time coincidences and because coordinate transformations do not alter such coincidences, the two mathematical solutions considered in the hole argument are observationally indistinguishable in principle. \footnote{Indeed, in a letter to Michelle Besso of January 3rd, 1916,~\cite{EinsteinCollectedPapers} Einstein gives essentially the same account of his error that he provided to Ehrenfest, and (according to the translation by
 Howard in Ref.~\cite{howard1992einstein})
  is explicit about the relevant notion being in-principle observability: 
 ``If, e.g., physical events were to be built up solely out of the movements of material points, then the meetings of the points, i.e., the points of intersection of their world lines, would be the only reality, i.e., observable in principle.''
  }

 From this, he concludes that the two solutions must be physically equivalent.  He is therefore making an inference from observational indiscernibility to ontological identity.
 
Overcoming the hole argument reaffirmed Einstein's confidence that the general theory of relativity should be generally covariant.  It proved to be a decisive step for it was not long afterwards that he formulated the final version of general relativity.  Therefore, we again find Leibniz's methodological principle to have played a critical role in Einstein's development of general relativity.

For the case of 
the point coincidence argument
 (unlike the cases of the induction experiment and the equivalence principle), the connection to Leibniz's philosophy has been noted before.
 Earman and Norton~\cite{earman1987price} point out that the space-time substantivalist must allow ``that there are distinct states of affairs which no possible observation could distinguish,'' and they note that this is the thesis that Leibniz emphatically argued against in his correspondence with Clarke.  Although they  do not explicitly describe the denial of this thesis as a version of Leibniz's principle of the identity of indiscernibles, preferring instead to refer to the particular instance of it required for the hole argument as 
 ``Leibniz equivalence'', it is clear that to deny ``that there are distinct states of affairs which no possible observation could distinguish'' is to affirm that in-principle empirical indistinguishability implies ontological identity, which is the version of Leibniz's principle of the identity of indiscernibles that I have been focussed on in this work.

Note, however, that Ref.~\cite{earman1987price} describes Einstein's temporary endorsement of the hole argument as ``tacitly eschewing Leibniz Equivalence''.  This assessment seems incorrect to me.
Such a conclusion would only be warranted if Einstein had in that period also endorsed the notion that the two scenarios under consideration were observationally equivalent.  However, this is {\em precisely} what Einstein missed prior to 1915.
Only when he recognized that the observational facts consisted solely of the point coincidences, so that the two scenarios {\em were} observationally equivalent after all, was he inclined to the opinion that they must be describing the {\em same} physical state of affairs and to conclude that demanding general covariance did not imply any nonuniqueness for the physical state of affairs inside the hole.
 Both before and after this realization, Einstein's views were consistent with the Leibnizian methodological principle.   In other words,  when one takes into account how he describes his mistake as a mistake regarding the observational discernibility of two scenarios, it becomes clear that his temporary endorsement of the hole argument does not indicate any tacit eschewment of the Leibnizian methodological principle, but rather constitutes further evidence of his commitment to it.





\subsection{A historical question}

An interesting historical question is whether one can  trace Einstein's endorsement of Leibniz's methodological principle back to Leibniz (either by direct or indirect influences),  or whether perhaps Einstein conceived of it independently.  I argue that, on balance, the historical evidence suggests the former. 

Einstein was intimately familiar with the debate on whether motion is absolute or relative, 
and therefore he was likely to have come across
 Leibniz's explicit use of the methodological principle is his argument against absolute space (such as the thought experiment wherein everything is shifted in space).  
Indeed, he refers explicitly to the debate in the foreword he wrote for
Max Jammer's book ``Concepts of Space''~\cite{jammer2013concepts} (in 1953, two years before his death).  Einstein begins by setting out the standard dichotomy between concepts of space, as: ``(a) space as positional quality of the world of material objects; (b) space as container of all material objects.''  He applauds Newton's endorsement of the concept (b) on the grounds that this was the only position that could be adequately justified in Newton's time:
\begin{quote}
The concept of space was enriched and complicated by Galileo and Newton, in that space must be introduced as the independent cause of the inertial behavior of bodies if one wishes to give the classical principle of inertia (and therewith the classical law of motion) an exact meaning. To have realized this fully and clearly is in my opinion one of Newton's greatest achievements. In contrast with Leibniz and Huygens, it was clear to Newton that the space concept (a) was not sufficient to serve as the foundation for the inertia principle and the law of motion. He came to this decision even though he actively shared the uneasiness which was the cause of the opposition of the other two: space is not only introduced as an independent thing apart from material objects, but also is assigned an absolute role in the whole causal structure of the theory. This role is absolute in the sense that space (as an inertial system) acts on all material objects, while these do not in turn exert any reaction on space. 
\end{quote}
He concludes, however, that Leibniz and Huygens' position was the correct one, and that even though they could not adequately justify it, their {\em intuitions} were well-founded:
\begin{quote}
The fruitfulness of Newton's system silenced these scruples for 
several centuries. Space of type (b) was generally accepted by 
scientists in the precise form of the inertial system, encompassing 
time as well. Today one would say about that memorable discussion: Newton's decision was, in the contemporary state of 
science, the only possible one, and particularly the only fruitful 
one. But the subsequent development of the problems, proceeding in a roundabout way which no one then could possibly foresee, has shown that the resistance of Leibniz and Huygens, intuitively well founded but supported by inadequate arguments, was actually justified. 
\end{quote}
This is consistent with the notion that Einstein endorsed Leibniz's use of the methodological principle as a criticism of position (a).  
The particular ``inadequacy'' in Leibniz's arguments to which Einstein here refers is likely a 
reference to
  the lack of a satisfactory account of the apparent empirical {\em discernibility} of unaccelerated and accelerated motions relative to absolute space (for instance, in the case of Newton's bucket experiment).
Indeed, Leibniz's response to the bucket experiment in the Leibniz-Clarke correspondence bears no resemblance to the response one sees to be the correct one in retrospect given relativity theory.
Indeed, it was not until Mach's analysis of the bucket experiment~\cite{mach1907science} that one has even the semblance of an adequate argument for position (a).


In addition to any direct influences of Leibniz on Einstein, there are many candidates for intellectual mediaries, individuals who were themselves directly influenced by Leibniz and who in turn influenced Einstein.   

Henri Poincar\'{e} is one such individual.  
In ``Science and Hypothesis'', published in 1902, 
Poincar\'{e} argued that because changes to the scale of the universe (i.e., global rescalings) led to no observable effects, they should not be considered physical.  Specifically, in defending the relativity of space~\cite{poincare2003science}, he mentions the following argument (which he attributes to Delboeuf):
\begin{quote}
Suppose that in one night all the dimensions of the universe became a thousand times larger.  The world will remain {\em similar} to itself, if we give the word {\em similitude} the meaning it has in the third book of Euclid.   Only, what was formerly a metre long will now measure a kilometre, and what was a millimetre long will become a metre.  the bed in which I went to sleep and my body itself will have grown in the same proportion.  when I wake in the morning what will be my feeling in face of such an astonishing transformation?  Well, I shall not notice anything at all.  The most exact measures will be incapable of revealing anything of this tremendous change, since the yard-measures I shall use will have varied in exactly the same proportions as the objects I shall attempt to measure.  In reality the change only exists for those who argue as if space were absolute.  If I have argued for a moment as they do, it was only in order to make it clearer that their view implies a contradiction.  In reality it would be better to say that as space is relative, nothing at all has happened, and that it is for that reason that we have noticed nothing.
\end{quote}


The example of the universe suffering a global rescaling overnight is so obviously similar to Leibniz's example of global translations mentioned earlier (and also of his other famous examples of global boosts and of flipping East into West), that it is hard to believe that Poincar\'{e} was not aware of the similiarity in the argumentation style.   Indeed, Poincar\'{e} was very familiar with  Leibniz's work. In 1880, for instance, the french edition of Leibniz's Monadology, prepared by \'{E}mile Boutroux, included a supplementary note at the end by Poincar\'{e} comparing Descartes' and Leibniz's conceptions of dynamics~\cite{Poincare1880Note}.  He also collaborated in preparing the international edition of the works of Leibniz~\cite{sep-poincare}.

Meanwhile, it is also known that Einstein was reading Poincar\'{e} in the years leading up to the development of the special theory of relativity\footnote{This occurred in the context of discussions of physics and philosophy with Maurice Solovine and Conrad Habicht (their self-proclaimed ``Olympia Academy'').  Specifically, in his reminiscences of Einstein, Solovine provided a list of the readings that served as the basis of their discussions, which included Poincar\'{e}'s ``Science and Hypothesis'' (Introduction to vol. 2 of Einstein's collected papers, Ref.~\cite{EinsteinCollectedPapers}, pp.~xxiv-xxv).
}
and so this is one avenue by which Einstein would have become acquainted (or {\em re}acquainted) with this Leibnizian argumentation style.
 %

\section{Response to criticisms of the principle}\label{ResponseToCriticisms}

A criticism of Leibniz's methodological principle that I sometimes hear is that to endorse it is to embrace a purely empiricist philosophy of science and hence a denial of realism.  

I have responded to this charge in another article~\cite{spekkens2015paradigm}, so I simply repeat the response here:
\begin{quote}
Such a principle does not force us to operationalism, the view that one should only seek to make claims about the outcomes of experiments.  For instance, if one didn't already know that the choice of gauge in classical electrodynamics made no difference to its empirical predictions, then discovery of this fact would, by the lights of the principle, lead one to renounce real status for the vector potential in favour of only the electric and magnetic field strengths.  It would not, however, justify a blanket rejection of \emph{any} form of microscopic reality.

As another example, consider the prisoners in Plato's cave who live out their lives learning about objects only through the shadows that they cast.  Suppose one of the prisoners strikes upon the idea that there is a third dimension, that objects have a three-dimensional shape, and that the patterns they see are just two-dimensional projections of this shape.  She has constructed a hidden variable model for the phenomena.  Suppose a second prisoner suggests a different hidden variable model, where in addition to the shape, each object has a property called colour which is completely irrelevant to the shadow that it casts.  The methodological principle dictates that because the colour property can be varied without empirical consequence, it must be rejected as unphysical. The shape, on the other hand, has explanatory power and the principle finds no fault with it.  Operationalism, of course, would not even entertain the possibility of such hidden variables.
\end{quote}


A variant of the criticism described above is that the Leibnizian principle implies a commitment to a particularly implausible doctrine of empiricism.  Specifically, it is suggested (at least in the context of the debate on the nature of space) that 
it amounts to nothing more than the verifiability criterion of meaningfulness of the logical positivists (see, e.g., Ref.~\cite{sklar1977space}, pp. 173-174), and consequently that the known deficiencies of the latter cut against it as well.

For instance, Earman and Norton~\cite{earman1987price} link the plausibility of the Leibnizian methodological principle to that of the doctrine of verificationism. 
Referring to the dilemma faced by spacetime substantivalists between (a) denying `Leibniz equivalence' (i.e., denying the Leibnizian methodological principle applied to the case of the hole argument) and (b) giving up spacetime substantivalism, they state that ``with the demise of the verifiability criterion of meaning, it is no longer unfashionable for them to escape the dilemma by simply allowing (a)''.   

But the two principles are not equivalent.  Although both make reference to observational indiscernibility of some pair of propositions, in the case of verificationism, such indiscernibility is taken as a sufficient condition for judging the propositions {\em meaningless}, while in the case of the Leibnizian methodological principle, such indiscernibility is merely taken as grounds for the implausibility of any ontological theory wherein  they are taken to describe physically distinct possibilities.



For instance, consider the examples of the application of the Leibnizian principle by Einstein that we have reviewed in this article.  Each example provides a criticism of an ontological theory that posits scenarios that are physically distinct but observationally indiscernible.  The ether theory of electromagnetism is the target of the criticism in the case of the induction experiment, while it is the Newtonian theory of gravitation in the case of the elevator thought experiment.  Does Einstein take the deficiency of these ontological theories to be that they are {\em meaningless}?  No. He criticizes them on the grounds that they fail to provide a satisfactory explanation of the relevant observational equivalence. \footnote{Note, furthermore that in light of Einstein's heavy use of the Leibnizian methodological principle, the claim that this principle is equivalent to the doctrine of verificationism is in tension with the fact that positivism cannot do justice to Einstein's philosophy of science. }



The Leibnizian methodological principle is 
best understood as an inference to the best explanation. The best explanation of the fact that two scenarios are in-principle observationally indiscernible
 is that they are associated to the same physical state of affairs.  
For instance, in the case of his comparison of an accelerating frame of reference and a frame fixed relative to a gravitational field, recall what Einstein says about explaining their empirical indiscernibility by their physical identity: the latter supposition makes the former fact (specifically, the observation of the equal acceleration of all bodies in a gravitational field) ``a matter of course''.

In Poincare\'{e}'s use of the Leibnizian methodological principle to argue for the unphysicality of global rescalings of space (discussed above),
it is even more explicit that he takes the principle  to be justified (in part at least)
 as an inference to the best explanation: ``as space is relative, nothing at all has happened, and that {\em it is for that reason} that we have noticed nothing.'' (emphasis added).




Leibniz's justification of the principle of the ontological identity of empirical indiscernibles in terms of his principle of sufficient reason~\cite{LeibnizClarke} may also be viewed as a characterization of the former principle as an inference to the best explanation.
Specifically, this holds if one understands Leibniz's principle of sufficient reason as a commitment to the idea that good explanations do not contain arbitrary elements,  that is, elements for which no reason can be given for them being one way as opposed to another.  (We will pursue this idea further in Sec.~\ref{EinsteinOnHisMPs}.)


Some commentators suggest that the Leibnizian methodological principle, though useful as a means of selecting among competing ontological theories, should not be taken as a constraint on theory construction.  
 This is the position that is suggested by Maudlin's discussion of the Newton-Leibniz debate in Ref.~\cite{maudlin1993buckets}.  
 Although he 
opines that ``one should be made at least uncomfortable by the postulation of empirically inaccessible physical facts,'' and consequently that ``[other things being equal], one would prefer a theory without them'', he nonetheless grants that theories violating the Leibnizian methodological principle are still viable on the grounds that ``Man is not a measure of all things, and there is no reason to believe that all real properties must fall within the power of human observation.''

I do not find this argument
persuasive because the Leibnizian methodological principle  does not appeal to a parochial kind of empirical indiscernibility, judged relative to the particular in-born capabilities of humans or their particular technological capabilities at a given historical moment, but rather to the {\em in-principle} variety of empirical indiscernibility.  This variety of indiscernibility must be understood as 
 indiscernibility for {\em any} system that might be considered an agent within the universe.  This is because,
  as Deutsch has argued persuasively~\cite{deutsch2011beginning}, the only in-principle limits to human capabilities
are the limits imposed by physics\footnote{His argument proceeds by noting that an ``in-principle human capability'' includes what could be achieved in a distant future with the aid of arbitrarily sophisticated technology.}, and therefore the only limits on our capabilities are the limits on the capabilities of {\em any} system embedded in the universe and subject to its physical laws.

Note, furthermore, that the in-principle variety of empirical indiscernibility is meaningful even in the absence of any agents whatsoever.  It is perfectly meaningful to assert that two scenarios {\em could not} in principle be discerned empirically (by the lights of some ontological theory) even in some early or late epoch of the universe where there were no agents about.  This is because an ontological theory specifies the limits on the discerning capacities of any system that could exist in the universe by its lights,  regardless of whether there are in fact any agents existing.


Finally, note that espousing Leibniz's methodological principle is not equivalent to denying the possibility of the underdetermination of physical theory by empirical evidence.  Rather, it is a denial of a {\em particular kind} of underdetermination.  If, within a given ontological theory, there is an element that can be varied without leading to any variation in observable phenomena (even in principle), then this is an underdetermination of {\em features} of that ontological theory by the empirical evidence, and it is {\em this} sort of underdetermination that is considered unacceptable 
 by the Leibnizian methodological principle.



\section{Prior art on the methodological principles at play in Einstein's work}


In this section, I will consider what a few previous authors have said concerning the methodological principles underlying Einstein's thought, and to what extent such claims are in agreement with the one made in this article. 


\subsection{The work of Elie Zahar}




 Elie Zahar, in attempting to articulate the methodological principles to which Einstein was committed (he termed them {\em heuristic} principles), described one such principle as follows~\cite{Zahar1976}:
\begin{quote}
All observationally revealed symmetries signify fundamental symmetries at the ontological level.  Hence the heuristic rule: replace any theory which does not explain symmetrical observational situations as the manifestation of deeper symmetries [...].
\end{quote}

If one interprets ``observationally revealed symmetries'' as invariances of the observations under putative changes to the ontological state,
then this statement is essentially equivalent in content to the statement of the Leibnizian methodological principle that I provided earlier in this article.  
This interpretation of Zahar's statement, therefore, supports our claim about Einstein's methodological commitments.  Note, however, that Zahar does not make explicit the connection with the philosophy of Leibniz.


\subsection{The work of Julian Barbour}

Julian Barbour has previously argued in favour of the thesis that Leibnizian ideas underlie 
Einstein's work. 
In particular, he has argued that one can understand many of Einstein's methodological commitments as instances of Leibniz's {\em principle of sufficient reason}.
For instance, he writes~\cite{barbour2003deep}:
\begin{quote}
If you read through Einstein's papers in which he battled his way to the creation of his general theory of relativity, you will see that the spur that kept him going was, in fact, the principle of sufficient reason. Indeed, he carried on directly from where Leibniz was forced by his untimely death to leave the issue. As Einstein never ceased to point out, Newton's use of absolute space was tantamount, in modern terms, to the introduction of distinguished frames of reference (for the formulation of the laws of nature) under conditions in which it was completely impossible to find any reason why they should be distinguished. Einstein found this to be an affront to the principle of sufficient reason, and was therefore led to say that no such distinguished frames of reference can exist, or, rather, that all conceivable frames must be equally good for the formulation of the laws of nature. This was his principle of general relativity---and what a harvest it eventually yielded.
That, I think, is enough justification for taking Leibniz seriously.
\end{quote}

As noted in Sec.~\ref{ResponseToCriticisms}, one can understand Leibniz's principle of the identity of indiscernibles (qua constraint on theory construction) as a particular instance of his principle of sufficient reason.
    Insofar as the hole and point-coincidence arguments and the arguments from the induction experiment and the equivalence principle are instances of the former, they can also be understood as instances of the latter.  In this sense, I agree with Barbour's assessment that Einstein was motivated to uphold Leibniz's principle of sufficient reason.  
  
  
Nonetheless, in my view, Barbour's summary does not fully capture
 the precise role that the principle of sufficient reason plays in Einstein's work.
Einstein only appeals to the  principle of sufficient reason, I claim, as a way of justifying the plausibility of the principle of the ontological identity of empirical indiscernibles.
 Specifically, he appeals to it in justifying the notion that good explanations do not invoke arbitrary elements, that is, 
  elements for which no reason can be given why they are one way rather than another.  
 But his endorsement of the principle of relativity --- in particular, the hypothesis that the laws of physics should take the same form in an accelerated frame as they do in an inertial frame with a gravitational field --- {\em cannot} be explained as merely a consequence of his dissatisfaction with theories containing arbitrary elements, and so the principle of sufficient reason does not {\em by itself} explain Einstein's endorsement of the sameness of the laws in the two frames.
   Rather, it is this principle {\em supplemented with the observed fact of the empirical indiscernibility of phenomena in the two frames} that Einstein took to be the justification for assuming the sameness of the ontological laws that operate in the two frames.

In other words, I take Barbour's summary of Einstein's reasoning to be incomplete insofar as it leaves out the important role of empirical observations (as opposed to merely rationalist considerations).  Once their role is acknowledged, in the manner I have just outlined, it becomes clear that  the particular form of the principle of sufficient reason that is at play is the principle of the ontological identity of empirical indiscernibles.

  
\subsection{The work of Don Howard}

Don Howard has argued~\cite{howard1992einstein} that a central aspect of Einstein's methodology is what he calls the {\em Eindeutigkeit} principle, which he takes Einstein to have picked up
 from the work of Joseph Petzoldt.\footnote{The german term ``Eindeutigkeit'' can be translated to ``unambiguity'', although Howard seems to prefer the synonymous but more obscure term ``univocalness'' in most of his work.}
Howard attempts to articulate the principle in various different ways in his article.
One such articulation appears to me to be cognate with Leibniz's principle of sufficient reason\footnote{Howard does make the connection to Leibniz's philosophy insofar as he notes that the {\em Eindeutigkeit} principle is reminiscent of the Leibnizian conception of mathematics.}:
\begin{quote}
Petzoldt gives the example of a stationary ball on a flat, horizontal, frictionless surface being struck, straight on, by a moving ball.  In what direction will the stationary ball recoil?  The only possibility is for it to recoil in the same direction in which the incident ball was moving, because for any other conceivable direction, say ten degrees to one side of the incident ball's path, one can find another direction, thus  ten degrees to the other side, that would be equally justified. 
\end{quote}
The language of ``equally justified options'' here is what is reminiscent of Leibniz's description of the lack of a sufficient reason to prefer one option over another.


The Leibnizian methodological principle defended in this article is a normative principle for the manner in which ontological theories ought to {\em secure explanations} of empirical phenomena, and it constrains empirical observations only 
insofar as it constrains the ontological theories one should be willing to consider as viable options.  Specifically, it asserts that the only viable options are those ontological theories
  that are not part of a slate of alternatives that are ontologically distinct but empirically equivalent.
In Howard's example of the application of the  {\em Eindeutigkeit} principle to the collision experiment, the principle is also used to argue
 in favour of the ontological theory that is not part a slate of alternatives
  (by eliminating all theories wherein the ball is deflected by some nontrivial angle on the grounds that each such theory is part of a pair of theories with opposite deflection angles).
But the slate of alternatives being considered in the collision experiment is a set of alternative ontological theories that lead to {\em different} empirical phenomena (because the angle of deflection of the ball is observable).


In other words, even if the  {\em Eindeutigkeit} principle can be understood as an instance of Leibniz's principle of sufficient reason, it is clearly not a restatement of the principle of the ontological identity of empirical indiscernibles insofar as it is not used to rule out ontological theories on the grounds of their having some element that can be varied without any empirical consequences.  
In my view, therefore, Howard's reading of Einstein's use of Leibniz, like Barbour's,
 does not do justice to the role of the empirical phenomena in Einstein's arguments.  The discussion in the next section will provide further support for this conclusion. 

\section{Einstein's own account of his methodological commitments}\label{EinsteinOnHisMPs}


To what extent did Einstein himself articulate the Leibnizian methodological principle when describing his own commitments? This is the question to which we turn in this section,

In his Autobiographical notes~\cite{einstein1970autobiographical},
 Einstein at one point sets out to ``say something general'' about ``the points of view from which physical theories may be analyzed critically at all.''  
He distinguishes two of these.
``The first point of view is obvious: the theory must not contradict empirical facts.''\footnote{\label{DuhemQuine}He qualifies this dictum by noting that its application is
 ``delicate'' on account of the fact that ``it is often, perhaps even always, possible to retain a general theoretical foundation by adapting it to the facts by means of artificial additional assumptions'', an endorsement of the thesis espoused by Duhem~\cite{duhem1991aim} and later by Quine~\cite{van1976two}. }
    ``The second point of view is not concerned with the relationship to the observations but with the premises of the theory itself, with what may briefly but vaguely be characterized as the `naturalness' or `logical simplicity' of the premises (the basic concepts and the relations between these).''  
 Regarding the distinction between the two points of view, he states: ``The second point of view may briefly be characterized as concerned with the `inner perfection' of the theory, whereas the first point of view refers to the `external confirmation.'\;{''}  
 
He mentions several methodological commitments that fall under the umbrella of the second point of view.   For example, he endorses a preference for theories that have greater specificity of predictions: ``among theories with equally `simple' foundations, that one is to be taken as superior which most sharply delimits the otherwise feasible qualities of systems (i.e., contains the most specific claims).''\footnote{From a modern perspective, it is natural to gloss this statement as a preference for theories that are more falsifiable relative to competitors which are less specific.  It is worth remembering that Karl Popper's falsifiability criterion was in part inspired by the example of Einstein's revolutionary work in physics.}

But of the commitments he articulates, the one that is of most relevance to the thesis of the present work is the following~\cite{einstein1970autobiographical}:
\begin{quote}
The following I reckon as also belonging to the ``inner perfection'' of a theory.  We prize a theory more highly if, from the logical standpoint, it does not involve an arbitrary choice among theories that are equivalent and possess analogous structures.
\end{quote}


I will refer to this as Einstein's {\em Wir sch\"atzen} statement.\footnote{This is the german phrase corresponding to ``We prize'' in the Schilpp translation of the Autobiographical notes~\cite{einstein1970autobiographical}.}

Some points to note.  Einstein is here articulating a criterion for theory-preference that is {\em not} simply the criterion of empirical adequacy because he explicitly situates the criterion under the second point of view, which pertains not to ``external confirmation'' but ``internal perfection''.    Thus when he compares on the one hand 
\begin{itemize}
\item[(i)] a theory that is one of a slate of theories that  are ``equivalent and possess analogous structures'',
\end{itemize}
to, on the other hand,
\begin{itemize}
\item[(ii)] a theory that is not of this sort, 
\end{itemize}
the two alternatives under consideration cannot be distinguished in terms of empirical adequacy.  
 

Einstein states that we ought to prefer option (ii), the theory that is {\em not} selected as one among a slate of theories that are ``equivalent'' and ``possess analogous structures.''  

Here, ``equivalent'' obviously cannot be interpreted as precise identity of the theories under consideration because then it would be impossible for their structures to be merely analogous and there would be no choice to make among them (not even an arbitrary one).
It must be a weaker notion of equivalence.  It must be strong enough, however, to imply empirical equivalence, for otherwise it would be possible in principle to distinguish the theories on the slate in terms of their empirical adequacy and Einstein has explicitly stipulated that this criterion is not to be situated under the ``external confirmation'' of a theory but rather its ``internal perfection''
  So it is reasonable to interpret the notion of equivalence at play here as one that implies empirical equivalence.  Indeed, it is reasonable to interpret it simply {\em as} empirical equivalence. 

How shall we conceive the property of 
``possessing analogous structures''?  
If the meaning of ``equivalence'' of the theories in the slate is indeed empirical equivalence,  then the structures to which Einstein is referring cannot be purely empirical structures because if they were, they would not have been merely {\em analogous} in these structures but {\em identical}.
 Thus the structures in question must refer to nonempirical aspects of the theory, that is, 
 physical or ontological aspects.


Under this reading, Einstein is asking us to consider a slate of theories that are empirically equivalent but that posit non-identical (though analogous) ontological structures.  And he is asserting that each theory in such a slate is made implausible by the fact that one cannot choose a single theory among them without this choice being arbitrary.  
It seems appropriate, therefore, to consider Einstein's statement here as an endorsement of the ontological identity of empirical indisernibles.

A comment of Einstein's shortly following the {\em Wir sch\"atzen} statement 
 provides further support for such a reading.
Einstein brings up his methodological commitments to set the stage for his critique of mechanics as the basis of physics.  After discussing critiques based on empirical adequacy, he turns to the second, ``interior'' point of view.  He then makes the following remark~\cite{einstein1970autobiographical}:
\begin{quote}
In today's state of science, i.e., after the abandonment of the mechanical foundation, such a critique retains only a methodological relevance.   But such a critique is well suited to show the type of argumentation that, in the selection of theories in the future, will have to play an ever greater role the more the basic concepts and axioms are removed from what is directly observable, so that the confrontation of the implication of theory by the facts becomes constantly more difficult and more drawn out. 
\end{quote}
What is pertinent here is his reference to the degree by which ``the basic concepts and axioms are removed from what is directly observable.''
This provides further evidence that, in the description of his methodology,  Einstein is explicitly concerned with the problem of bridging between the empirical evidence and the ontological theory.

Following this remark, he  gives the critique of mechanics from the second point of view.  
His critique focusses on the preference given to inertial systems over other rigid coordinate systems in the Newtonian conception of physics.  
Criticizing this preference fits the mold of the {\em Wir sch\"atzen} statement
 by virtue of the fact that among all rigid coordinate systems, picking out the inertial systems as preferred
   is an arbitrary choice.  Here, Einstein is appealing to the fact it is possible to express the laws of physics in the same manner relative to all rigid coordinate systems, and that the empirical predictions are the same for all choices.
This is just an appeal to the Leibnizian methodological principle of the sort that we've seen already in our discussion of the equivalence principle. 
The only difference is that Einstein does not merely focus 
 on the empirical indistinguishability of an accelerating frame in free space and an inertial frame with a uniform gravitational field, but on the empirical indistinguishability of any rigid coordinate system with some corresponding inertial coordinate system and gravitational field. 

So the critique of mechanics which he provides as an example of his methodological principle at work is aptly characterized as an argument based on the principle of the ontological identity of empirical indiscernibles.

Finally, it is worth asking how Einstein means to judge in-principle empirical indiscernibility.  In my statement of the Leibnizian methodological principle, I have explicitly stated that it ought 
to be judged {\em by the lights of the ontological theory}.  Such an attitude towards how indiscernibility is to be judged is consistent with Einstein's own statements about his philosophy of science, in particular, his famous dictum that ``what is observable is determined by the theory"
  and his general endorsement of Duhem's thesis 
   that all observations are theory-laden (see footnote~\ref{DuhemQuine} and Ref.~\cite{howard1990einstein}).

In summary, Einstein's own account of one of his methodological principles seems to conform to the Leibnizian methodological principle as I have articulated it here.   
Certainly, the {\em Wir sch\"atzen} statement
 reads quite well as an attempt to describe what is common to the arguments of his that we reviewed in this article: the induction experiment, the equivalence principle, and the hole and point-coincidence arguments.
Nonetheless,
one can certainly imagine Einstein having provided a {\em better} summary of the Leibnizian methodological principle if this is what he intended.  
Indeed, it is particularly notable that he did not add qualifiers that the equivalence he was  referring to was empirical, nor that the ``analogous structures'' were ontological.

 
It is also worth noting, however, that the {\em Wir sch\"atzen} statement is followed immediately by Einstein apologizing
for not providing more detail~\cite{einstein1970autobiographical}: ``I shall not attempt to excuse the lack of precision of the assertions contained in the last two paragraphs on the grounds of insufficient space at my disposal; I must confess herewith that I cannot at this point, and perhaps not at all, replace these hints by more precise definitions.  I believe, however, that a sharper formulation would be possible.''

\color{black}

\section{Discussion}

\subsection{The relevance of the Leibnizian methodological principle in physics today}

Thus far, I have sought to make the case for the Leibnizian methodological principle 
 based on its historical track record.   
Given the heavy use that Einstein made of the principle in formulating the conceptual pillars on which the theory of Relativity was built,
its credentials are already seen to be quite impressive.  In light of this, I believe that physicists would do well to form the habit of considering what the principle entails. 
It involves a particular application of operational facts in the service of realism which is rather different from the methodological commitments of most physicists.  
I therefore turn now to
 the question of what the Leinibizian methodological principle can do for physics today.

For one, various contentious issues in the foundations of physics can be usefully reexamined from the perspective of this principle.
As an example, the deBroglie-Bohm interpretation of quantum theory
 has been criticized on the grounds that it requires the existence of an absolute rest frame, even though, by the lights of the theory, this frame can never be determined by any observation. 
   Anyone who endorses this criticism
    has intuitions in line with the Leibnizian methodological principle.\footnote{I thank Matt Leifer for the example.} There are in fact many aspects of pilot-wave theories that are underdetermined by the empirical data, as noted in Ref.~\cite{spekkens2015paradigm}, and, therefore, by the lights of this principle, each of these constitutes a reason for being skeptical of this interpretational approach.  (A proponent of the deBroglie-Bohm interpretation has some recourse, however, as discussed in the next section.)
The principle is also pertinent to the study of 
noncontextuality in quantum theory.  In Ref.~\cite{spekkens2005contextuality}, I proposed a generalization of the notion of noncontextuality espoused by Kochen and Specker~\cite{KS}
to one that is applicable to any operational theory and any type of experimental procedure therein.  This principle, understood as applying to all types of experimental procedures and termed {\em generalized noncontextuality}, is motivated in a manner that is precisely analogous to the Leibnizian methodological principle (see, e.g., Appendix A of Ref.~\cite{mazurek2016experimental}).
There is a subtlety however: to properly understand the principle of generalized noncontextuality as a special case of the Leibnizian methodological principle, it is necessary to reconceive the latter at the level of epistemology.
That is, rather than conceiving of it as an inference from the indiscernibility of empirical facts to the identity of ontological facts, one must conceive of it as an inference from the indiscernibility of {\em states of knowledge about} empirical facts to the identity of {\em states of knowledge about} ontological facts.
  A detailed discussion is beyond the scope of this work
   (see, however, Ref.~\cite{SchmidSelbySpekkens}). 

I pause to address a potential argument {\em against} the Leibnizian methodological principle that arises from its connection to noncontextuality.
Specifically, sceptics of the principle could take the fact that quantum theory {\em does not} admit of a generalized-noncontextual ontological model as evidence {\em against} the principle of generalized noncontextuality and hence also as evidence against the Leibnizian methodological principle.  To see that the argument need not lead one to having scepticism of the principle,
 it suffices to note that generalized noncontextuality is not the only assumption of the no-go theorem and consequently that one can preserve generalized noncontextuality (and hence the Leibnizian methodological principle)  by giving up a different assumption.
  In my opinion, this is in fact the right attitude to take towards the no-go result.  Specifically, I believe that it is the framework of ontological models that must be abandoned, and that it is a fruitful research programme to seek an alternative to this framework that provides causal explanations of quantum statistics while strictly respecting the Leibnizian methodological principle~\cite{spekkens2015paradigm,leifer2013towards, wood2015lesson, allen2017quantum}.  

More generally, the 
principle may well reveal important deficiencies in our current approach to theory construction in physics.
In Ref.~\cite{spekkens2015paradigm}, I argued that a commitment to  it implies
 that the distinction between kinematics and dynamics---which is central to theory-construction in physics today---is unphysical and that only a kind of union of the two (somehow capturing causal relations alone) can have ontological significance. 
This is the sort of novel conception of the ontological account of phenomena that I alluded to above as a potential alternative to the standard framework of ontological models.
  At the very least, such a conception provides  
 a fresh perspective  on the problem of understanding the ontological claims of quantum theory, a project which I firmly believe is critical to the progress of modern physics.  
  
 \subsection{A middle ground in the debate between realists and empiricists}

Researchers investigating the foundations of physics tend to polarize into two camps, corresponding to the divide between realist and empiricist philosophies of science.
 This is particularly true in the foundations of quantum theory, where a kind of schism has developed between the most enthusiastic proponents of each of the two communities.
In this section, I will argue that to endorse the Leibnizian methodological principle is to stake out a middle ground between empiricism and realism.
Specifically, I will argue that 
an endorsement of the principle removes the most problematic features of each camp's position and therefore serves to 
draw researchers on both ends of this ideological divide towards the centre

A criticism that can be levelled 
against realists is that the ontological theories they entertain
 are underdetermined by the empirical evidence.
For concreteness, consider a realist
 who is enthused by the deBroglie-Bohm interpretation of quantum theory.
An operationalist may criticize such a researcher's position (rightly in my view)  by noting 
that there are many possible theories that fit the mold of pilot-wave interpretations, and that these different theories vary in their ontological commitments but are not distinguished by the empirical evidence~\cite{spekkens2015paradigm}.   

Suppose such a realist decides to endorse the Leibnizian methodological principle. Call them a ``Leibnizian Bohmian''.  
If they do so, then
 they must agree that this type of underdetermination is indeed problematic insofar as it explicitly violates the principle.  Nonetheless, they can still defend a research program that espouses the deBroglie-Bohm interpretation on the grounds that {\em future empirical evidence} could distinguish between the different versions of the theory.  Indeed, to stay true to the principle, they must grant that were future empirical evidence {\em not} to distinguish these versions, then this would be grounds for abandoning
the approach.  Furthermore, they should be motivated to 
 actively seek out such evidence.

It seems to me that Antony Valentini is a proponent of the deBroglie-Bohm interpretation of this type given that he has responded to the underdetermination criticism of pilot-wave theories 
 in precisely this fashion~\cite{valentini2004personal}.  Furthermore, the version of pilot-wave theories that he espouses~\cite{valentini1991signal1,valentini1991signal2,valentini2005dynamical} --- wherein there is the possibility of divergence between the initial probability distribution over the configuration space of the Bohmian particles and the modulus squared of the wavefunction --- is unlike the usual version insofar as it makes empirical predictions that are different from those of standard quantum theory.  
 It is therefore possible in principle to have differences of empirical predictions
    among instances of the theory that vary in their ontological commitments. Valentini's research program is to try and identify exotic physical
 scenarios that can reveal such differences, and consequently that could deliver a verdict on {\em which} pilot-wave theory in a slate of such theories 
 is the correct one. 
  Although I am not personally enthused by the prospects of this program, I prefer it to the standard version of the pilot-wave program precisely because it does not conflict
   with the Leibnizian methodological principle.

Conversely, a criticism that can be levelled against the empiricists is that 
their set of mental tools is too impoverished.  The idea is that in order to make progress in physics, 
one has no choice but to 
entertain ontological hypotheses, expressed with concepts that are not defined purely in terms of empirical phenomena (Einstein's ``free inventions of the human mind'').\footnote{Einstein himself made precisely this criticism against those empiricists who were skeptical of the reality of atoms (he names Ostwald and Mach in particular)~\cite{einstein1970autobiographical}: 
\begin{quote}The hostility of these scholars toward atomic theory can undoubtedly be traced back to their positivistic philosphical attitude.  This is an interesting example of the fact that even scholars of audacious spirit and fine instinct can be hindered in the interpretation of facts by philosophical prejudices.  The prejudice --- which has by no means disappeared --- consists in the belief that facts by themselves can and should yield scientific knowledge without free conceptual construction.
\end{quote}} 

An empiricist who endorses the Leibnizian methodological principle can address this criticism without conceding too much to the realist
by stipulating that ontological hypotheses should only be entertained when these are 
subservient to empiricist priorities.
Specifically, they can concede that it is
valuable to entertain an ontological hypothesis
but only
 if there is no element of that hypothesis that can be varied without empirical consequence.  
 
To use the example of the prisoners in Plato's cave mentioned earlier, the Leibnizian empiricist can retreat from a position of admitting {\em no} hidden variables to a position of admitting only those hidden variables that are analogous to the {\em shapes} of the objects casting the shadows on the cave wall while still rejecting as idle metaphysics any hidden variables that are analogous to the {\em colours} of the objects. 
 \subsection{Physics and philosophy} 
 
The question of which methodological principles might best promote scientific progress belongs primarily to the philosophy of science.  But the question is critical for the progress of physics.   Indeed, of all the assumptions that might differentiate a physicist from their peers, their methodological principles are the most important because they affect  the complexion of  {\em everything else that they assume} and the true worth of their labours will ultimately be determined by the veracity and fruitfulness of that starting point. 

I suspect that some physicists may 
 be sceptical of this idea because they imagine that physics can be pursued fruitfully without presuming {\em any} particular philosophical commitments.  I believe that such scepticism is fundamentally mistaken, 
 for reasons that have been articulated before.  Dennett expresses the rebuttal  to this view
  particularly well~\cite{dennett1996darwin}:
\begin{quote}
[...] there is no such thing as philosophy-free science. There is only science whose philosophical baggage is taken on board without examination. 
\end{quote}
Moreover, it is telling  that all the {\em truly} significant  revolutions in physics were accompanied by upheaval in the philosophy of science.  In other words: the physicists who were true revolutionaries could not afford to be 
philosophically na\"ive.

For any reader who needs further persuasion of the importance to physics of methodological principles in general (and the Leibnizian methodological principle in particular), I advise reflection on the question of why it was Einstein rather than Poincar{\'e} who discovered the special theory of relativity. 
 As Isaacson has elegantly put it in his biography of Einstein~\cite{isaacson2011einstein}: ``If this seems surprising [that it was Einstein, not Poincar{\'e}], it is because we underestimate the boldness of Einstein in stating the principle of relativity as an axiom and, by keeping our faith with it, changing our notion of space and time.''  I agree with this sentiment.  Indeed, a similar sort of answer could be given to the question of why it was Einstein who discovered the general theory of relativity rather than any other physicist who had come to learn of the special theory.  However, in answering either question, the principle to which I would point as the one constituting Einstein's central axiom
  is the
  Leibnizian methodological principle for theory construction,
   for the reasons that I have articulated in this article.    
    In my view, it was the fact that Einstein kept his faith in {\em this} principle
     that ultimately set him apart from his contemporaries in the account of history.  

\begin{acknowledgments}
Special thanks 
to Julian Barbour, who first
opened my eyes to the appeal of Leibnizian ideas, and to
 Doreen Fraser and David Schmid for their helpful comments on a draft of this manuscript.  Research at Perimeter Institute is supported by the Government of Canada through the Department of Innovation, Science and Economic Development Canada and by the Province of Ontario through the Ministry of Research, Innovation and Science.
\end{acknowledgments}



\bibliographystyle{unsrt}		
\bibliography{bib}

\begin{thebibliography}{10}

\bibitem{LeibnizClarke}
Gottfried Wilhelm~Freiherr von Leibniz and Samuel Clarke.
\newblock {\em The Leibniz-Clarke Correspondence: Together Wiith Extracts from
  Newton's Principia and Opticks}.
\newblock Manchester University Press, 1998.

\bibitem{spekkens2015paradigm}
Robert~W Spekkens.
\newblock The paradigm of kinematics and dynamics must yield to causal
  structure.
\newblock In Anthony Aguirre, Brendan Foster, and Zeeya Merali, editors, {\em
  Questioning the Foundations of Physics}, pages 5--16. Springer, 2015.

\bibitem{sklar1977space}
Lawrence Sklar.
\newblock {\em Space, time, and spacetime}, volume 164.
\newblock University of California Press, 1977.

\bibitem{einstein1905electrodynamics}
Albert Einstein.
\newblock On the electrodynamics of moving bodies.
\newblock {\em Annalen der Physik}, 17(891):50, 1905.

\bibitem{einstein1911influence}
Albert Einstein.
\newblock On the influence of gravitation on the propagation of light.
\newblock {\em Annalen der Physik}, 35(898-908):906, 1911.

\bibitem{einstein1907relativity}
Albert Einstein.
\newblock On the relativity principle and the conclusions drawn from it.
\newblock {\em Jahrbuch der Radioaktivitat und Elektronik}, 4:411--462, 1907.

\bibitem{einstein2005constructed}
Albert Einstein.
\newblock How {I} constructed the theory of relativity.
\newblock {\em Translated by M. Morikawa. Association of Asia Pacific Physical
  Societies (AAPPS) Bulletin}, 15(2):17--19, 2005.

\bibitem{einstein1916foundation}
Albert Einstein.
\newblock The foundation of the {G}eneral {T}heory of {R}elativity.
\newblock {\em Annalen Phys.}, 14:769--822, 1916.
\newblock English translation by Satyendra Nath Bose available at
  {en.wikisource.org}.

\bibitem{stachel1985einstein}
J~Stachel.
\newblock Einstein's search for general covariance, 1912-1915, paper read at
  the {N}inth {I}nternational {C}onference on {G}eneral {R}elativity and
  {G}ravitation, {J}ena 1980; published in {E}instein and the {H}istory of
  {G}eneral {R}elativity, {E}instein {S}tudies, vol. 1, eds. {D}. {H}oward and
  {J}. {S}tachel, 1985.

\bibitem{norton1984einstein}
John Norton.
\newblock How {E}instein found his field equations: 1912-1915.
\newblock {\em Historical studies in the physical sciences}, 14(2):253--316,
  1984.

\bibitem{stachel2014hole}
John Stachel.
\newblock The hole argument and some physical and philosophical implications.
\newblock {\em Living Reviews in Relativity}, 17(1):1, 2014.

\bibitem{NortonSEP}
John~D. Norton.
\newblock The hole argument.
\newblock In Edward~N. Zalta, editor, {\em The Stanford Encyclopedia of
  Philosophy}. Metaphysics Research Lab, Stanford University, summer 2018
  edition, 2018.

\bibitem{einstein1914foundations}
Albert Einstein.
\newblock On the foundations of the generalized theory of relativity and the
  theory of gravitation.
\newblock {\em Phys. Z.}, 15:176--180, 1914.

\bibitem{EinsteinCollectedPapers}
{A}lbert {E}instein.
\newblock {\em The {C}ollected {P}apers of {A}lbert {E}instein}.
\newblock Princeton University Press, 1987.

\bibitem{howard1992einstein}
Don Howard.
\newblock Einstein and eindeutigkeit: A neglected theme in the philosophical.
\newblock {\em Studies in the history of general relativity}, 3:154, 1992.

\bibitem{earman1987price}
John Earman and John Norton.
\newblock What price spacetime substantivalism? {T}he hole story.
\newblock {\em British Journal for the Philosophy of Science}, pages 515--525,
  1987.

\bibitem{jammer2013concepts}
Max Jammer.
\newblock {\em Concepts of space: the history of theories of space in physics:
  third edition}.
\newblock Courier Corporation, 2013.

\bibitem{mach1907science}
Ernst Mach.
\newblock {\em The science of mechanics: A critical and historical account of
  its development}.
\newblock Open court publishing Company, 1907.

\bibitem{poincare2003science}
Henri Poincar{\'e}.
\newblock {\em Science and method}.
\newblock Courier Corporation, [1914] 2003.

\bibitem{Poincare1880Note}
Henri Poincar\'{e}.
\newblock {\em Note sure les principes de la m\'{e}canique dans {D}es{C}artes
  et dans {L}eibnitz}.
\newblock 1880.

\bibitem{sep-poincare}
Gerhard Heinzmann and David Stump.
\newblock Henri {P}oincar\'{e}.
\newblock In Edward~N. Zalta, editor, {\em The Stanford Encyclopedia of
  Philosophy}. Metaphysics Research Lab, Stanford University, winter 2017
  edition, 2017.

\bibitem{maudlin1993buckets}
Tim Maudlin.
\newblock Buckets of water and waves of space: Why spacetime is probably a
  substance.
\newblock {\em Philosophy of science}, 60(2):183--203, 1993.

\bibitem{deutsch2011beginning}
David Deutsch.
\newblock {\em The beginning of infinity: Explanations that transform the
  world}.
\newblock Penguin UK, 2011.

\bibitem{Zahar1976}
Elie Zahar.
\newblock Why did {E}instein's programme supersede {L}orentz's?
\newblock In Colin Howson, editor, {\em Method and appraisal in the physical
  sciences: {T}he critical background to modern science, 1800-1905}. Cambridge
  University Press, 1976.

\bibitem{barbour2003deep}
Julian Barbour.
\newblock The deep and suggestive principles of {L}eibnizian philosophy.
\newblock {\em The Harvard Review of Philosophy}, 11(1):45--58, 2003.

\bibitem{einstein1970autobiographical}
Albert Einstein.
\newblock Autobiographical notes in {A}lbert {E}instein: Philosopher-scientist.
\newblock {\em Library of Living Philosophers, Cambridge UP, London}, 7:33,
  1970.

\bibitem{duhem1991aim}
Pierre Maurice~Marie Duhem.
\newblock {\em The aim and structure of physical theory}, volume~13.
\newblock Princeton University Press, 1991.

\bibitem{van1976two}
Willard van Orman~Quine.
\newblock Two dogmas of empiricism.
\newblock In {\em Can Theories be Refuted?}, pages 41--64. Springer, 1976.

\bibitem{howard1990einstein}
Don Howard.
\newblock Einstein and {D}uhem.
\newblock {\em Synthese}, 83(3):363--384, 1990.

\bibitem{spekkens2005contextuality}
Robert~W Spekkens.
\newblock Contextuality for preparations, transformations, and unsharp
  measurements.
\newblock {\em Physical Review A}, 71(5):052108, 2005.

\bibitem{KS}
Simon Kochen and Ernst Specker.
\newblock The problem of hidden variables in quantum mechanics.
\newblock volume~17, page~59, 1967.

\bibitem{mazurek2016experimental}
Michael~D Mazurek, Matthew~F Pusey, Ravi Kunjwal, Kevin~J Resch, and Robert~W
  Spekkens.
\newblock An experimental test of noncontextuality without unphysical
  idealizations.
\newblock {\em Nature Communications}, 7:11780, 2016.

\bibitem{SchmidSelbySpekkens}
David Schmid, John Selby, and Robert~W. Spekkens.
\newblock {\em In preparation}, 2019.

\bibitem{leifer2013towards}
Matthew~S Leifer and Robert~W Spekkens.
\newblock Towards a formulation of quantum theory as a causally neutral theory
  of {B}ayesian inference.
\newblock {\em Physical Review A}, 88(5):052130, 2013.

\bibitem{wood2015lesson}
Christopher~J Wood and Robert~W Spekkens.
\newblock The lesson of causal discovery algorithms for quantum correlations:
  Causal explanations of {B}ell-inequality violations require fine-tuning.
\newblock {\em New Journal of Physics}, 17(3):033002, 2015.

\bibitem{allen2017quantum}
John-Mark~A Allen, Jonathan Barrett, Dominic~C Horsman, Ciar{\'a}n~M Lee, and
  Robert~W Spekkens.
\newblock Quantum common causes and quantum causal models.
\newblock {\em Physical Review X}, 7(3):031021, 2017.

\bibitem{valentini2004personal}
Antony Valentini.
\newblock Personal communication.
\newblock 2004.

\bibitem{valentini1991signal1}
Antony Valentini.
\newblock Signal-locality, uncertainty, and the subquantum {H}-theorem. {I}.
\newblock {\em Physics Letters A}, 156(1-2):5--11, 1991.

\bibitem{valentini1991signal2}
Antony Valentini.
\newblock Signal-locality, uncertainty, and the subquantum {H}-theorem. {II}.
\newblock {\em Physics Letters A}, 158(1-2):1--8, 1991.

\bibitem{valentini2005dynamical}
Antony Valentini and Hans Westman.
\newblock Dynamical origin of quantum probabilities.
\newblock {\em Proceedings of the Royal Society A: Mathematical, Physical and
  Engineering Sciences}, 461(2053):253--272, 2005.

\bibitem{dennett1996darwin}
Daniel~C Dennett.
\newblock {\em Darwin's Dangerous Idea: Evolution and the Meanings of Life}.
\newblock Simon and Schuster, 1996.

\bibitem{isaacson2011einstein}
Walter Isaacson.
\newblock {\em Einstein: His life and universe}.
\newblock Simon \& Schuster, New York, USA, 2007.

\end{thebibliography}

\end{document}